\documentclass{sf2a-conf2015}
\usepackage{graphicx}
\usepackage{hyperref}
\usepackage{epstopdf}

\begin{document}

\TitreGlobal{SF2A 2015}


\title{The GSO Data Centre}

\runningtitle{The GSO Data Centre}

\author{F. Paletou}\address{Universit\'e Paul Sabatier, OMP-IRAP, CNRS,
  14 ave. E. Belin, F--31400 Toulouse, France}
\author{J.-M. Glorian$^1$}
\author{V. G\'enot$^1$}
\author{A. Rouillard$^1$}
\author{P. Petit$^1$}
\author{A. Palacios}\address{Universit\'e de Montpellier 2, LUPM,
  CNRS, Place E. Bataillon, F--34095 Montpellier, France}
\author{E. Caux$^1$}
\author{V. Wakelam}\address{Universit\'e de Bordeaux, LAB, CNRS, F--33270
  Floirac, France}

\setcounter{page}{237}


\maketitle


\begin{abstract}
  Hereafter we describe the activities of the \emph{Grand Sud-Ouest}
  Data Centre operated for INSU (CNRS) by the OMP--IRAP and the
  \emph{Universit\'e Paul Sabatier} in Toulouse, in a collaboration
  with the OASU--LAB in Bordeaux and OREME--LUPM in Montpellier.
\end{abstract}



\section{Introduction}

The GSO Data Centre (hereafter OV--GSO) was officially set-up in 2013,
after approval by INSU of CNRS. Its role is to support more specific
and so-called ``reference services'' which provide dedicated and
communautary services, in relation with relevant astrophysical
data. OV--GSO also promotes and encourages the deployment of virtual
observatory (VO) techniques, at the regional level.

The actual distribution of regional data centres at the national
level can be seen in Fig.\,(1). OV--GSO covers all the \emph{open and
  ``science ready'' data}, and VO-oriented activities of Bordeaux
(LAB), Montpellier (LUPM) and Toulouse (IRAP) laboratories for
astrophysics.

\begin{figure}[t!]
 \centering
 \includegraphics[width=0.8\textwidth,clip]{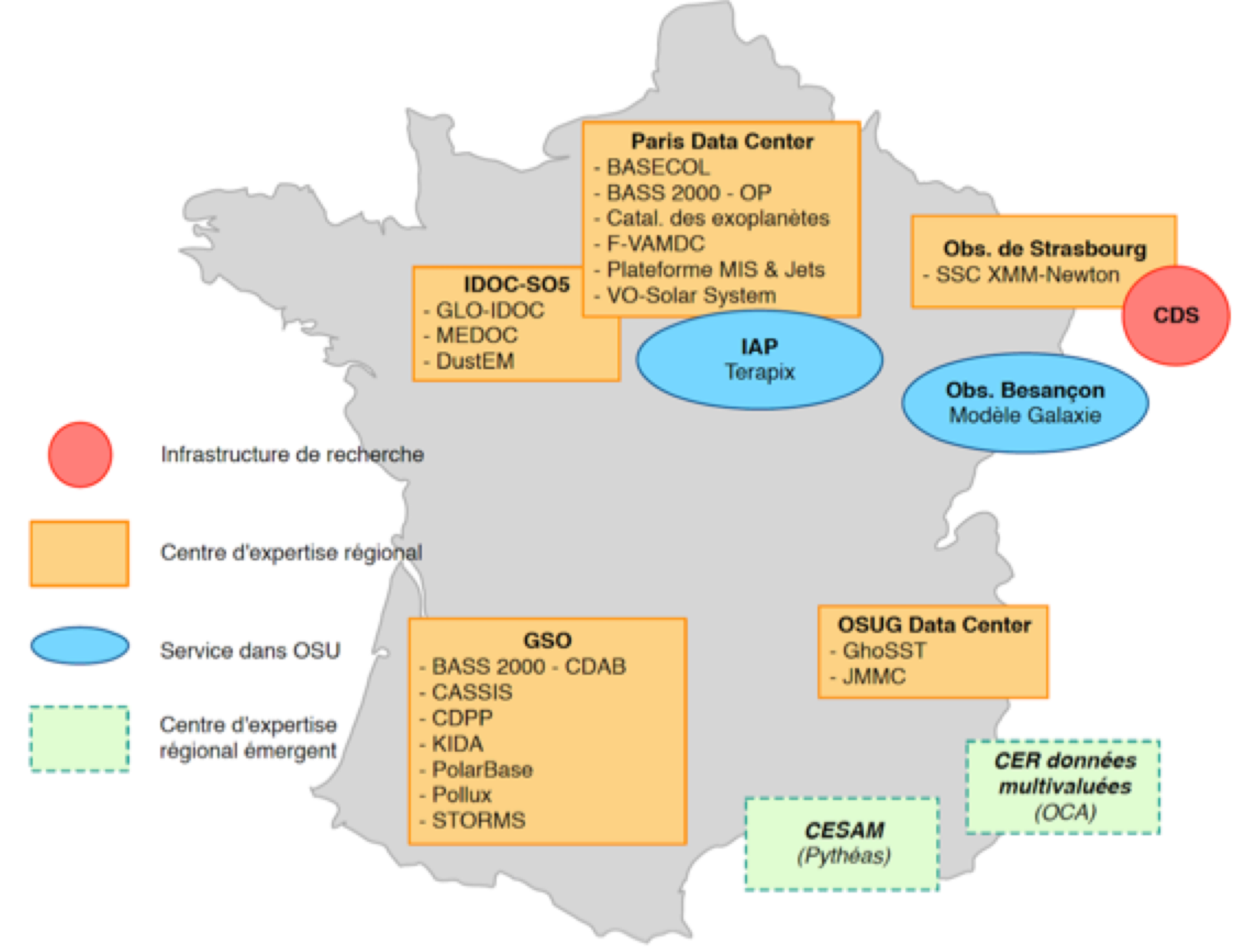}      
 \caption{Actual map of astronomical data services in France. OV--GSO
   covers the whole south-west area i.e., the (open) data-oriented
   activities of Bordeaux (LAB), Toulouse (IRAP) and Montpellier
   (LUPM) astrophysics laboratories.}
  \label{fig1}
\end{figure}

\section{Reference services}

\subsection{Bass\,2000}

{\sc Bass\,2000} was originally set-up in 1996 for the archival and
diffusion of all the solar data collected from ground-based
observatories having a national participation. At this time, it was
essentially giving direct support to the Th\'emis solar telescope
located at the \emph{Observatorio del Teide}, Tenerife (Spain).

The progressive decline of the Th\'emis community, and the evolution
of data management at the era of virtual observatories makes that this
service will now focus on the only support to the ground-based
instrument under responsability of the \emph{Observatoire
  Midi--Pyr\'en\'ees}, that is the CLIMSO set of coronagraphs and
narrow-band solar (disk) imagers located at the summit of \emph{Pic du
  Midi de Bigorre}.
 
\subsection{CDPP}

The CDPP is the national centre of expertise concerning
terrestrial and planetary plasma data. It was created in 1998 by
CNRS/INSU and the French space agency CNES. It assures the long term
preservation of data obtained primarily from instruments built using
French resources, and renders them readily accessible and exploitable
by the international community. The CDPP also provides services to
enable on-line data analysis (AMDA, see {\tt amda.cdpp.eu}), and 3D
data visualization in context (3DView, see {\tt 3dview.cdpp.eu}).

The  CDPP also plays an important role in the development of
interoperability standards (see e.g., G\'enot et al. 2014).
 
\subsection{STORMS}

STORMS, which stands for \emph{Solar Terrestrial ObseRvations and
  Modeling Service}, is a new public service providing tools and data to
perform studies in heliophysics and space weather, and to study and
model the influence of solar activity on the geospace environment, as
well as on planets or any other solar system bodies (comets,
asteroids or spacecrafts).

The main tool it provides so far, {\tt propagationtool.cdpp.eu}, was
jointly developed with CDPP. It is meant for the tracking of solar
storms, streams and energetic particles in the heliosphere.
 
\subsection{PolarBase}

{\sc PolarBase} was officially opened to the public in 2013. This
service distributes high resolution optical stellar spectra from the
Espadons@CFHT and Narval@TBL spectropolarimeters.

Reduced spectra, in various Stokes parameters, are delivered to the
community, as well as standardizely extracted polarized signatures. A
complete description of the database can be found in Petit et al. (2014).
 
\subsection{POLLUX}

{\sc Pollux} is a stellar spectra database proposing access to
\emph{theoretical} data. For that purpose, high resolution synthetic
spectra and spectral energy distributions have been computed using the
best available models of atmosphere (CMFGEN, ATLAS and MARCS),
performant spectral synthesis codes (CMF\_FLUX, SYNSPEC and
TURBOSPECTRUM) and atomic linelists from VALD database and specific
molecular linelists for cool stars. Spectral types from O to M are
represented for a large set of fundamental parameters: T$_{\rm eff}$, log$g$,
[Fe/H], and specific abundances (Palacios et al. 2010).
 
\subsection{CASSIS}

CASSIS started in 2005. It provides an interactive spectrum analyser
that was originally proposed for the scientific exploitation of
(far-infrared and submillimetric) data from the Herschel Space
Observatory. CASSIS allows to visualize observed or synthetic spectra,
together with a line identification tool. It can also predict spectra
which may be observed by any (single-dish, so far)
telescope. Comparison between observations and synthetic spectra
(e.g., from Radex) is also possible with the same tool.

CASSIS is now evolving towards a multi-purpose spectral analysis
tool, operating beyond its initial range of application.
 
\subsection{KIDA}

KIDA is a database of kinetic data of interest for astrochemical
(interstellar medium and planetary atmospheres) studies. In addition
to the available referenced data, KIDA provides recommendations over a
number of important reactions. Chemists and physicists can also add
their own data to the database.

KIDA also distributes a code, named Nahoon, to study the
time-dependent gas-phase chemistry of 0D and 1D interstellar
sources. Details about the KIDA database can be found in Wakelam et
al. (2012).

\begin{figure}[t!]
 \centering
 \includegraphics[width=0.88\textwidth,clip]{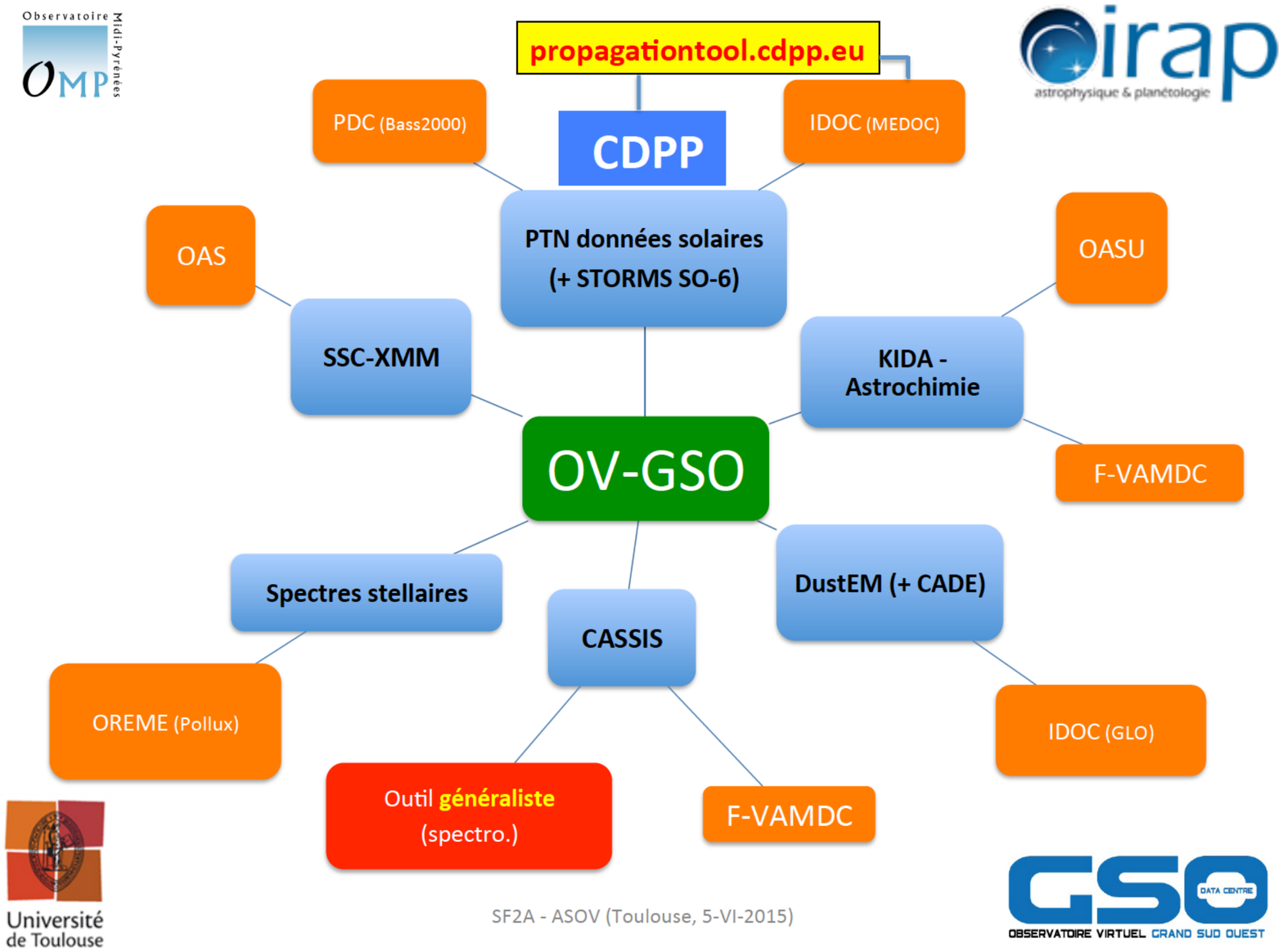}      
 \caption{Connections between OV--GSO and other services and
   institutes at the national level.}
  \label{fig2}
\end{figure}

\section{Management and operations}

As of 2015, operations of OV--GSO involve about 10 technical (IT)
personnel, and about 25 scientists in the
Bordeaux-Toulouse-Montpellier area. The typical annual budget of the
data centre is about 50 kEur. Obviously one of the major task of
OV--GSO is to guarantee the continuity of all services, and therefore
regular upgrades of all hardware devices are made. Another fundamental
task of the data centre is to plan and to increment regularly our data
storage capability, in an homogeneous way.

We also regularly contribute to the various virtual observatories
communities, both at the national and international levels (e.g.,
IVOA). This concerns our reccurent participation to the bi-yearly
so-called {\sc Interop}  meetings of the IVOA, as well as propositions of
tutorials (e.g., {\sc SpecFlow} at {\tt euro-vo.org} scientific tutorials
page, or Paletou \& Zolotukhin 2014).

Locally, we set-up a monthly dedicated seminar, oriented towards the
use and implementation of Virtual Observatory standards and
protocols. We are also involved in discussions about more general data
issues, at the level of the \emph{Universit\'e de Toulouse}.  Our
various activities can be followed at {\tt ov-gso.irap.omp.eu}

\section{Perspectives}

Figure (2) finally summarizes the various interactions between OV--GSO
and other services and institutes, at the national level (and as of
2015).

We are however already supporting several ongoing projects which
should transform into new reference services in a near future. It
consists in providing data and tools for the study of emissions of the
extended sky ({\tt cade.irap.omp.eu}), and a so-called Multi Frequency
Follow-up Inventory Service ({\tt muffins.irap.omp.eu}). Projects are
also associated with the data management of the {\sc Muse}
integral-field spectrograph at VLT, as well as high-energy
astrophysical data, from spaceborne missions such as XMM (see e.g.,
{\tt xmmssc.irap.omp.eu}) or Fermi, to
the ground-based CTA facility (see e.g., {\tt cta.irap.omp.eu}). An
involvement in the data and VO-oriented activities of the space
mission {\sc Euclid} is also planned.

\begin{acknowledgements}
  OV--GSO is supported by {\sc Cnrs/Insu}, the \emph{Universit\'e Paul
    Sabatier, Observatoire Midi--Pyr\'en\'ees, Toulouse}, and the
  \emph{Action Sp\'ecifique ``Observatoire Virtuel''} of {\sc Cnrs}.
\end{acknowledgements}



%
\end{document}